\documentclass[preprint2,longabstract]{aastex}

\renewcommand{\b}[1]{\mbox{\boldmath $#1$}}

\def\cal#1{{\cal #1}}

\def\m@th{\mathsurround=0pt}
\def\n@space{\nulldelimiterspace=0pt \m@th}
\def\biggg#1{{\mbox{$\left#1\vbox to 20.5pt{}\right.\n@space$}}}

\def\beginenum{\begin{enumerate}}
\def\endenum{\end{enumerate}}
\def\bitem{\begin{itemize}}
\def\eitem{\end{itemize}}
\def\bray{\begin{array}}
\def\eray{\end{array}}
\def\begindoc{\begin{document}}
\def\enddoc{\end{document}}
\def\bq{\begin{equation}}
\def\eq{\end{equation}}
\def\bqy{\begin{eqnarray}}
\def\eqy{\end{eqnarray}}
\def\bqyn{\begin{eqnarray*}}
\def\eqyn{\end{eqnarray*}}
\def\bc{\begin{center}}
\def\ec{\end{center}}
\def\bfll{\begin{flushleft}}
\def\efll{\end{flushleft}}
\def\bflr{\begin{flushright}}
\def\eflr{\end{flushright}}
\newcommand{\Avec}{\mbox{\boldmath $A$}}
\newcommand{\Bvec}{\mbox{\boldmath $B$}}
\newcommand{\Evec}{\mbox{\boldmath $E$}}
\newcommand{\Fvec}{\mbox{\boldmath $F$}}
\newcommand{\Gvec}{\mbox{\boldmath $G$}}
\newcommand{\Rvec}{\mbox{\boldmath $R$}}
\newcommand{\Uvec}{\mbox{\boldmath $U$}}
\newcommand{\Vvec}{\mbox{\boldmath $V$}}
\newcommand{\evec}{\mbox{\boldmath $e$}}
\newcommand{\jvec}{\mbox{\boldmath $j$}}
\newcommand{\kvec}{\mbox{\boldmath $k$}}
\newcommand{\nvec}{\mbox{\boldmath $n$}}
\newcommand{\uvec}{\mbox{\boldmath $u$}}
\newcommand{\vvec}{\mbox{\boldmath $v$}}
\newcommand{\wvec}{\mbox{\boldmath $w$}}
\newcommand{\xvec}{\mbox{\boldmath $x$}}
\newcommand{\omegavec}{\mbox{\boldmath $\omega$}}
\newcommand{\Omegavec}{\mbox{\boldmath $\Omega$}}

\usepackage{spr-astr-addons}    

\usepackage{color}

\RequirePackage{color}
\def\imagei{\centerline{\color[gray]{.75}\rule{\hsize}{4pc}}}%
\def\imageii{\centerline{\color[gray]{.75}\rule{4pc}{4pc}}}%

\newcommand{\vdag}{(v)^\dagger}
\newcommand{\emaila}{authors@email.com}

\begin{document}
%
\title{Mechanism for flow generation/acceleration in dense degenerate stellar atmospheres}

\shorttitle{<Flow generation in degenerate stellar atmospheres>}
\shortauthors{<Barnaveli \& Shatashvili >}

\author{ A.A. Barnaveli\altaffilmark{1,2}}
\altaffiltext{1}{Department of Physics, Faculty of Exact \&
Natural Sciences, Javakhishvili Tbilisi State University,
Tbilisi 0179, Georgia}
\altaffiltext{2}{Department of Physics,
Faculty of Science,  Utrecht University, 3508 TC Utrecht, The
Netherlands}
\author{ N.L. Shatashvili\altaffilmark{1,3}}
\altaffiltext{1}{Department of Physics, Faculty of Exact \&
Natural Sciences, Javakhishvili Tbilisi State University,
Tbilisi 0179, Georgia} \altaffiltext{3}{Andronikashvili Institute
of Physics, TSU, Tbilisi 0177, Georgia}

\begin{abstract}
The mechanism for flow generation in dense degenerate stellar
atmospheres is suggested when the electron gas is degenerate and
ions are assumed to be classical. It is shown, that there is a
catastrophe in such system -- fast flows are generated due to
magneto-fluid coupling near the surface. Distance over which
acceleration appears is determined by the strength of gravity and
degeneracy parameter. Application of this mechanism for White
Dwarfs' atmospheres is examined and appropriate physical parameter
range for flow generation/acceleration is found; possibility of
the super-Alfv\'enic flow generation is shown; the simultaneous
possibility of flow acceleration and magnetic field amplification
for specific boundary conditions is explored; in some cases
initial background flow can be accelerated 100 and more times
leading to transient jet formation while the Magnetic field
amplification is less strong.

\end{abstract}


\maketitle


\section{Introduction}

Several recent studies were devoted to mechanisms of multi-scale
equilibrium structure formations in astrophysical plasmas
\citep{bib:Shiraishi,bib:iqbal-3,bib:pino,bib:Multi-B} based on the so called
Beltrami-Bernoulli (BB) class of equilibria model. Such structures
were found applicable to study the heating of solar atmosphere as
well as the problems of the large-scale magnetic and velocity
field generation \citep{bib:mmns-1,bib:ymois,bib:mnsy,bib:msms}; \\
\cite{bib:osym}. Most investigations were limited to non-degenerate
plasmas so that the constituent particles were assumed to obey the
classical Maxwell-Boltzman statistics. Latest studies
\citep{bib:BSM_deg,bib:SMB_multi} examined the possibility of the
transformation of equilibrium BB states for the highly dense and
degenerate plasmas applicable to Compact Star conditions (mean
inter-particle distance is smaller than the de Broglie thermal
wavelength) so that the particle energy distribution was dictated
by Fermi-Dirac statistics [see \citep{bib:degenerate} and references
therein]. Such highly dense/degenerate plasmas are also found in various
astrophysical and cosmological environments as well as in the
laboratories devoted to inertial confinement and high energy
density physics \citep{bib:Dunne-1,bib:Yanovski,bib:Dunne-2,bib:Dunne-3}.

In the theory of BB equilibria plasma flows play crucial role.
Interestingly enough, recent observations prove that flows are
widely present in solar atmosphere
\citep{bib:flows-1,bib:Trace,bib:flows-3}; \\
\citep{bib:flows-2,bib:flows-4};\\
\citep{bib:flows-5,bib:Aschwanden-2};\\
\citep{bib:Hansteen,bib:flows-jets};\\
\citep{bib:flows-new,bib:flows-spicules}
and their role in the dynamics and heating of multi-scale
complex-structure solar corona is already well appreciated
although the theoretical investigations were performed
\citep{bib:mmns-1,bib:ymois,bib:mnsy,bib:msms} long before
the observational evidence of their existence.
Flows are found crucial in astrophysical disks [see e.g.
\citep{bib:vinod,bib:zanni,bib:SY-DJ}; \\
\citep{bib:Chagelishvili,bib:bodo}] and their corona, in inter- and
extra-galactic environments. If so, then finding the steady flow
generation/acceleration mechanisms in astrophysical conditions is
of the utmost importance [see e.g.
\citep{bib:Goodman,bib:Uchida,bib:Koutchmy,bib:mnsy}].

\bigskip

Compact astrophysical objects like white and brown dwarfs,
neutron stars, magnetars with characteristic electron number
densities within $(10^{26}\div 10^{32})\,cm^{-3}$ are the
natural habitats for dense/degenerate matter \citep{bib:Chandra1,bib:Chandra2};\\
\citep{bib:Compact-1,bib:Begelman};\\
\citep{bib:Compact-2,bib:White};\\
\citep{bib:Sturok-3,bib:Beloborodov};\\
\citep{bib:Shukla-1,bib:Shukla-2}. \\
When a star collapses, and cools down, density of lighter elements
increases affecting the total pressure/enthalpy of unit fluid
element; degenerate electrons provide the dominant pressure, while
the only significant source of energy is the reservoir of thermal
energy in the nearly classical ideal gas of ions (see the review
\citep{bib:winget} and references therein). Cooling process seems to
be sensitive to outer layers/atmosphere composition, structure and
their conditions. Up to now there doesn't exist a precise model of
atmospheres of White Dwarfs (WDs) although recent studies show
that a significant fraction of White Dwarfs are found to be
magnetic with typical fields strengths $<1\,KG$. At the same time,
massive and cool White Dwarfs are found with much higher fields
detected [see \citep[and references therein]{bib:kepler}].

Magnetic WDs are considered to be stellar remnants featuring
global magnetic structures with field strengths within $1\,kG \div
1000\,MG$ \citep{bib:tremblay}. They account for a significant part of
WD population \citep{bib:hmfwd,bib:schmidt};\\
\citep{bib:Kawka}. Most of these objects are high-field magnetic WDs,
with field strengths $B> 1\,MG$, and a distribution of magnetic
field strengths that appears to peak around $B>20\,MG$
\citep{bib:schmidt,bib:kulebi}. The origin of magnetic WDs remains
elusive. At the same time, according to \citep[and references
therein]{bib:tremblay}, "the magnetic fields (at the $10MG$ level)
could have an influence on the structure of corresponding WD
models at best only in the outermost $0.5\% $ of the radius"; they
demonstrated that convective energy transfer is seriously impeded
by magnetic fields when plasma-$\beta $ becomes $<1$; the critical
field strength that inhibits convection in the photosphere of WDs
is in the range $B = (1-50) kG$, which is much smaller than the
typical $1-1000MG$ field strengths observed in magnetic white
dwarfs, implying that these objects have radiative atmospheres.
Studying the cooling process of high-field magnetic WDs ($B
\gtrsim 10 MG$), authors found that the full suppression of
convection has no effect on cooling rates until the effective
temperature reaches a value of around $5500\,K$. In this context
it is interesting that recent investigations \citep[and references
therein]{bib:DAZ} have uncovered several cool, magnetic, polluted
hydrogen atmosphere WDs (DAs). Using the calcium lines they
determined a surface averaged magnetic field of $B_S = 0.335 \pm
0.003 MG $ with a velocity of $v = 19.8 \pm 1.7 km/s $. Using
$H_{\alpha}$ they determined $B_S = 0.331 \pm 0.004 MG $ with a
velocity of $v = 23.9 \pm 2.9 km/s $ -- the magnetic field
measurements and velocities were consistent within uncertainties.
It was also found that the incidence of magnetism in old,
polluted WDs (DAZ) significantly exceeds what is found in the
general WD population suggesting a hypothetical link between a
crowded planetary system and magnetic field generation. It was
shown that WD stars with such surface temperatures that convection
zones develop, seems to show stronger magnetic fields than hotter
stars; the mean mass of magnetic stars seems to be on average
larger than the mean mass of nonmagnetic WD stars.

Pure-hydrogen-atmosphere white dwarfs convection zone study
of \citep{bib:tremblayConv} with the 3D simulations restricted to its upper part
suggests (through hydrodynamical calculations) that the
entropy of the upflows does not change significantly from the
bottom of the convection zone to regions immediately below the
photosphere; to calibrate 1D envelopes authors relied on this
asymptotic entropy value, characteristic of the deep and
adiabatically stratified layers. The calibration encompasses the convective
hydrogen-line (DA) WDs in the effective temperature range
$6000\ll T_{eff}(K)\ll 15000$ and the surface gravity range
$7.0\ll log (g) \ll 9.0$.
In typical cases for DA WDs, they found convective
velocities to be of the order of $V_{z,rms}\sim 1\,km/s$ at the base
of the convection zone reaching maximum value of $6\,km/s$.

According to \citep{bib:hollands} the exact fraction of WDs which have
strong ($B \gtrsim 3$\,MG) surface magnetic fields is of order
$10\%$ . Moreover, as stated above, the magnetic WDs are
systematically more massive \citep{bib:ferrario}. This could be
partially explained by the core dynamo-generated fields; the
related field strengths are right within the observed distribution
of WD surface fields \citep{bib:ferrario}. It is argued that many
(perhaps the majority of) WDs could contain strong ($B \gtrsim
10^6$\,G) magnetic fields which are confined within the stellar
interior and not detectable at the surface even as they cool. This
is because the WD cooling timescale is shorter than its magnetic
diffusion timescale $\equiv 10^{11}$ years \citep{bib:cumming}. These
fields may have very important implications for WD evolution, and
for the outcome of WD mergers.

Thus, it is very important to find the mechanism for the origin
and evolution of surface magnetic fields as well as studying the
multi-structure dynamics of WD atmospheres in connection to both
magnetic and velocity fields generation, heating/cooling; to
uncover the effects of magneto-fluid couplings in outer layers of
accreting stars since the dynamical evolution of their convective
envelopes may define the final structure of their interior as well
as of atmospheres. Knowledge of well-investigated similar
processes in solar and sun-like star atmospheres can serve as a
guidance for such investigations -- the latest observations have
demonstrated that the solar corona as well as the chromosphere is
a highly dynamic arena replete with multiple-scale spatiotemporal
structures \citep{bib:Aschwanden}.


Plasma flows were found to complement the abilities of the
magnetic field in the creation of richness observed in Stellar
atmospheres \citep{bib:mnsy,bib:msms}. The most obvious process for flow
generation could be the conversion of magnetic and/or thermal
energy to plasma kinetic energy. Magnetically driven transient
(but sudden) flow generation mechanisms permeate the literature.
For a more quiescent pathway Mahajan et al 2002 applied the
Bernoulli mechanism which converts thermal energy into kinetic
energy, or to the general magneto-fluid rearrangement of a
relatively constant kinetic energy, i.e., going from an initial
high-density/low-velocity state to a low-density/high-velocity
state. In above studies the double--Beltrami-Bernoulli states
accessible to a two-fluid system \citep{bib:MY,bib:iqbal-1} provided the necessary
framework.

As shown in \citep{bib:BSM_deg};\\
\citep{bib:SMB_multi}, when the star contracts its outer layers keep
the multi-structure character although density in structures
becomes defined by electron degeneracy pressure. In this respect
it would be extremely important to examine the flow generation
process predicted in these papers as well as the accompanying
phenomena for the understanding of concrete compact objects' outer
layers dynamics with degenerate plasma, specifically, for the
evolution picture of magnetic White Dwarfs (WD). In present paper
we present the results of detailed study of magneto-fluid coupling
process in two-fluid plasma with degenerate electrons exploring
the acceleration mechanism, highlighting the differences with
classical case and finding the appropriate physical parameter
ranges for flow generation/acceleration, transient jet formation,
magnetic field amplification.

\section{Model Equations for Flow generation in Dense Degenerate Stellar Atmospheres}

In \citep{bib:BSM_deg} it was developed a simplest model in which the
effect of quantum degeneracy on the nature of the BB class of
equilibrium states was illustrated. Emphasizing the quantum
degeneracy effects, it was chosen a model hypothetical system
(relevant to specific aspects of a WD) of a two-species neutral
plasma with non-degenerate non relativistic ions, and degenerate
relativistic electrons embedded in a magnetic field. It was
assumed that, despite the relativistic mass increase, the electron
fluid vorticity is negligible compared to the electron cyclotron
frequency (such a situation may pertain, for example, in the
pre-WD state of star evolution, or in the dynamical magnetic WD
atmosphere). Study of the degenerate electron (positron) inertia
effects on Beltrami States in dense neutral plasmas constituted
the scope of \citep{bib:SMB_multi} where the generation mechanisms of
dynamical large (meso) scale were demonstrated; this scale, though
present in a classical non-degenerate plasma, turns out to be
degeneracy dependent and could be vastly different from its
classical counterpart.

Restricting ourselves to steady state considerations our approach
will be based on a straightforward application of the developed
magnetofluid theory of \citep{bib:MY,bib:mmns-1,bib:ymois,bib:SMB_multi} assuming
that at some height of magnetic WDs surface there exist the fully
ionized magnetized plasma structures such that the
quasi-equilibrium two-fluid model of \\
\citep{bib:BSM_deg} will capture the essential physics of flow or/and
magnetic field amplification. Corresponding equilibrium state
equations are as follows [stating that for the non-relativistic
ions and inertialess electrons there are two independent Beltrami
conditions aligning ion and electron generalized vorticities along
their respective velocities]:
\begin{equation}
{\bf b} = a\,N\,\left[{\bf V} - \frac{1}{N}\nabla \times {\bf b}
\right] \ , \label{DB1}
\end{equation}
\begin{equation}
{\bf b}+\nabla \times {\bf V} = d\,N\,{\bf V} \ , \label{DB2}
\end{equation}
where \ ${\bf b}=e{\bf B}/m_ic$ \ and it was assumed, that
electron and proton laboratory-frame densities are nearly equal -
$N_e\simeq N_i=N$ \ [rest-frame density \ $n_{e,i} =
N_{e,i}/\gamma ({\b V}_{e,i})$ \ with \ $\gamma ({\b V}_{e,i})$ \
being a Lorentz factor for electrons/ions]; here \ $a$ \ and \ $d$
\ are dimensionless constants related to the two invariants: the
magnetic helicity \ $h_1=\int{({\bf A}\cdot {\bf b})\,d^3x}$ \ and
the generalized helicity \ $h_2=\int{({\bf A}+{\bf V})\cdot ({\bf
b}+\nabla \times {\bf V}) \,d^3x}$ \ of the system with \ ${\bf
A}$ \ being the dimensionless vector potential. Equations
(\ref{DB1}) and (\ref{DB2}) are written in terms of normalized one
fluid variables: the fluid velocity \ $\bf V$ \  and the current \
${\bf J}=\nabla \times {\bf b}$ \ (via Ampere's law) when the
electron and the ion speeds are given by ${\bf V}_e= {\bf
{V}}-(1/N)\nabla\times {\bf b}$, and \ ${\bf V}_i={\bf V}$ ,
respectively (the electrons, as stated above, are assumed to be
inertia less). Notice, that in this approximation the electron
vorticity is primarily magnetic \ (${\bf b}$) \ while the ion
vorticity has both kinematic and magnetic parts \ (${\bf b}+\nabla
\times {\bf V}$). Also, in above equations, the density is
normalized to $N_0$ (the corresponding rest-frame density is \
$n_0$); the magnetic field is normalized to some ambient $B_0$ ;
all velocities are measured in terms of the corresponding Alfv\'en
speed \ $V_A=B_0/\sqrt{4\pi N_0m_i}$ ; all lengths [times]  are
normalized to the skin depth \ $\lambda_i \ [{\lambda_i}/{V_A}]$ ,
where $\lambda_i = c/\omega_{pi}=c\,\sqrt{{m_i}/{4\pi N_0e^2}}$ .
We would like to comment here that the fundamental result of
vortex dynamics of any ideal fluid is that it implies a
topological invariant, the helicity (generaliced helicity), i.e.
such an invariance forbids the creation of vorticity (generalized
vorticity) from no-vorticity (no-generalized vorticity) state of
the two-fluid system \citep{bib:M-min,bib:RelVorticity,bib:pino}, \\
\citep{bib:curved}, and of more
general multi-species ideal plasma
\citep{bib:Multi-B}. The reason of such invariance lie deep in the
Hamiltonian structure governing the dynamics of the ideal fluid
\citep{bib:MY,bib:morrison}, \\
\citep{bib:arnold} and is independent of compressibility; however this
constraint gets broken for non-ideal multi-species plasmas. In
latter case all the corresponding helicities become dynamical,
hence, defining the fate of the system locally \\
\citep{bib:msms}.

To define an equilibrium state (the stationary solution of the
dynamical system) the Beltrami conditions (\ref{DB1}) and
(\ref{DB2})  shall be supplemented by the Bernoulli constraint
relating the density with the flow kinetic energy through
degeneracy pressure,  which reads as
\begin{equation}
\nabla \left(\beta_0\,{\rm{ln}}\,N + \mu_0\,\sqrt{1+P_F^2}\ \gamma
+ \frac{V^2}{2}\right) = 0 \ \label{Bernoulli}
\end{equation}
where $\beta_0$ is the ratio of the thermal pressure to the
magnetic pressure, \ $\mu_0=m_ec^2/m_iV_A^2$ \  and for the
electron fluid Lorentz factor \ $\gamma_e \simeq \gamma({\bf V})$
\ was used. Since the normalized Fermi momentum of electrons \
$P_F = p_F/m_e\,c=(NN_0/n_c\gamma)^{1/3}$ \ is a function of the
density \ $N$ (here $n_c=5.9\times 10^{29}\,cm^{-3}$  is the
critical number-density at which the Fermi momentum equals $m_e
c$), the system of equations
(\ref{DB1})-(\ref{DB2})-(\ref{Bernoulli}), together with the
automatically satisfied equilibrium continuity equation \ [$\nabla
\cdot ({N{\bf V}})=0$] \ and the divergence free condition for
magnetic field \ [$\nabla \cdot {\bf b}=0$] , form a fully
specified equilibrium -- a complete system to determine \
$\bf{N}$, ${\bf V}$ and $\bf{b}$. Notice, that for significant \
$P_F(N)$ [relevant to WD's conditions] in (\ref{Bernoulli}) the
degeneracy pressure can be much bigger than the thermal pressure
(measured by $\beta_0$) and, hence, the cold (low $\beta_0$)
degenerate electron gas can sustain a qualitatively new state:
there exists a nontrivial Double BB equilibrium at zero/low
temperatures \citep{bib:BSM_deg} [in contrast to the classical
zero-beta plasmas, where only the relatively trivial, single
Beltrami states are accessible \citep{bib:ymois}]. We will show below
that, when gravity effects are also taken into account, exactly
this interesting departure from classical situation makes it
possible to couple the magnetic and velocity fields in degenerate
dense and cold WD stars outer-layers / atmospheres discussed in
the introduction.

Bernoulli condition (\ref{Bernoulli}) introduces a qualitatively
new player in the equilibrium balance -- the spatial variation in
electron degeneracy energy (proportional to $\mu_0 =
m_ec^2/m_iV_A^2$ ) could increase / decrease the fluid kinetic
energy (measured by $V^2$) or plasma $\beta_0$ in the
corresponding region of star's outer layer / atmosphere converting
Fermi energy to the kinetic/thermal energy or vice versa; it could
also forge a re-adjustment of kinetic energy from a
high-density/low-velocity plasma to a low-density/high-velocity
plasma. Related energy transformations, mediated through classical
gravity in solar atmosphere, were discussed in Mahajan et al
(2002,2006). While for the special class of magnetic WDs study of
\citep{bib:BSM_deg} predicted that the electron degeneracy effects can
be both strong and lead to the anti-correlation between density
and flow speeds -- the generated flow gets locally
super-Alfv\'enic in contradistinction to non-degenerate, thermal
pressure dominated Solar Atmosphere plasma (with local plasma
$\beta < 1$) for which the maximal velocity due to the
magneto-Bernoulli mechanism was found to be only locally
sub-Alfv\'enic.

\bigskip

In present paper we go further and construct the detailed
solutions of (\ref{DB1}-\ref{Bernoulli}) with gravity taken into
account for concrete parameters relevant to magnetic WDs to show
the explicit effects of degeneracy on two-fluid BB structures when
star contracts and cools down. First we rewrite these equations
with inclusion of classical (Newtonian) gravity for nonrelatvistic
flows [$\gamma({\bf V})\sim 1$ \ justified by observations] as
follows:
\begin{equation}
{\bf b} = a\,N\,\left[{\bf V} - \frac{\kappa}{N}\nabla \times {\bf
b} \right] \ , \label{DB3}
\end{equation}
\begin{equation}
{\bf b}+ \kappa\,\nabla \times {\bf V} = d\,N\,{\bf V} \ , \label{DB4}
\end{equation}
\begin{equation}
\nabla \left(\beta_0\,{\rm{ln}}\,N + \mu_0\,\sqrt{1+P_F^2}\
-\frac{R_A}{R} + \frac{V^2}{2}\right) = 0 \ \label{Bernoulli2}
\end{equation}
where $R$ is a radial distance from the center of WD normalized to
its radius \ $R_{W}\ [\sim (0.008 - 0.02)\ R_{\odot}]$ and
$R_A=GM_{W}/R_{W}V_A^2$ (here $G$ is the gravitational constant,
$M_W$ is a WD mass);  dimensionless parameter $\kappa =
\lambda_i/R_w$ .

For {\bf typical} cold magnetic WDs with the degenerate electron
densities $\sim (10^{25}-10^{29})\,cm^{-3}$; magnetic fields $\sim
(10^5-10^{9})\,G $, and temperatures \ $\sim (40000-6000)\,K $
(and, hence, with corresponding Alfv\'en speed\ $V_A\sim 
(10^4-10^6)\,cm/s$) yielding  $\beta_0 \sim (10^6-10^0)$ , {\bf
$R_A \ \sim (10^8 - 10^4) $} \ and \ $\mu_0 \sim (10^{10}-10^6)\gg
\beta_0$ , \ $\lambda_i \sim (10^{-5}-10^{-7})\,cm$ (being
rather small). Comparing the terms in (\ref{Bernoulli2}) for above parameters, one can
see the dominance of electron fluid degeneracy pressure in equilibrium balance.
Then, neglecting the first term related to ion fluid pressure, \ Eqs. (\ref{DB3}), (\ref{DB4}),
(\ref{Bernoulli2}) can be easily manipulated to obtain:
\begin{equation}
\frac{\kappa^2 }{N}\nabla \times \nabla \times {\bf V} + \kappa
\nabla \times \left[ \left( \frac{1}{aN}-d \right)N{\bf V} \right]
+ \left[ 1-\frac{d}{a} \right]{\bf V} = 0  , \label{DB5}
\end{equation}
\begin{equation}
N=\left[ \frac{ \left( \left[ \frac{R_A }{R} -
\frac{R_A}{R_0}\right]-\left[ \frac{{\bf V}^2}{2} - \frac{{\bf
V_0}^2}{2}\right] + \alpha \mu_0^2 
\right)^2}{a_0^2\mu_0^2} - \frac{1}{a_0^2} \right]^{3/2} ,
\label{density}
\end{equation}
with \ $a_0=(n_0/n_c)^{1/3}$ \ and \ $\alpha =
\sqrt{1+a_0^2N_0^{2/3}}$ ; here subscript ''0'' is used for the
height from stellar surface where the boundary conditions are
applied. These equations provide us with a closed system of four
equations in four variables \ $(N, \, {\bf V})$ . Once this is
solved with appropriate boundary conditions for magnetic WDs, one
can invoke (\ref{DB4}) to calculate ${\bf b}$ . The reader can
find the solution for the similar mathematical problem relevant to
the non-degenerate (solar atmosphere) case in \citep{bib:mnsy}.

We must emphasize, that these time-independent equations are not
suitable for studying the heating/cooling processes, the latter
requires a fully time-dependent treatment; main thrust of this
communication is to uncover the mechanisms that create/accelerate
flows and amplify the magnetic fields in outer layers of WDs.
Since \ $P_F = a_0\,N^{1/3}$ \ is a function of Fermi energy (and,
hence, a function of density), Bernoulli equation
(\ref{Bernoulli2}) relates density with flow kinetic energy and
classical gravity [Eq. (\ref{density})] in a qualitatively
different way from classical case. Also, since for most class of
magnetic WDs outer-layer/atmosphere conditions the degeneracy
pressure is larger than the thermal pressure in
Eq.(\ref{Bernoulli2}), the new results are eventually expected in
this very limit of star evolution while its contruction/accretion.
At the same time it is evident that due to the same reason the
influence of discussed nonlinear coupling process on the cooling
rate of compact object is automatically eliminated; possibly this
proves the findings of \citep{bib:tremblay} about the non-affected
cooling process relevant for most of cold magnetic WDs.

\section{Simulation Results for Flow generation in Dense Degenerate Stellar Atmospheres}

To explore the mechanisms for flow and magnetic field generation /
amplification in dense degenerate stellar atmospheres we have
carried out a one-dimensional extensive simulation experiments
(the relevant dimension being the height ''Z'' from the center of
WD; $Z_0 = R_W + \Delta R $ is the surface at which the boundary
conditions are applied) of the coupled nonlinear system of
equations (\ref{DB5} - \ref{density}) for a variety of boundary
conditions. The results are presented in Figures 1-8. These are
the plots of different physical characteristic quantities as
functions of height. For all our runs the boundary conditions
$|{\bf b}_0|=1$ and $V_0 = a^{-1}V_{A,0}$ (with $a\sim d ; \
V_{x,0}=V_{y,0}=V_{z,0}$ in Cartesian Geometry) were imposed (see
equations (\ref{DB3}) and (\ref{DB4})) and due to the limitations
of the code the values chosen for \ $\kappa $ \ (measure of the
strength of two-fluid Hall currents) were larger than their actual
values for WD atmospheres but, as seen from the illustrative
results, the physical tendency of the nonlinear couplings are well
seen even for such assumptions. We have chosen two different DB
parameters ( $d \sim a$ ) limits to start with various initial
set-ups; e.g. (i) choice of \ $d \sim a \gg 1$ \ means that our
initial flow is locally sub-Alfv\'enic (strongly magnetic WD outer
layer) while (ii) the opposite choice of \ $d \sim a \ll 1$ \
would mean that we are dealing with the super-Alfv\'enic flow to
start with (weakly magnetic WD outer layer). Notice, that since
densities of WDs' outer layer / atmosphere degenerate plasma are
very high the Alfv\'en speeds can be rather low locally and we can
deal with both extreme cases for initial setups -- even the weak
initial flow can become super-Alfv\'enic locally. It is
important to emphasize here that even the ion skin depth is very
small, its contribution may become very crucial [see for solar
case \citep{bib:mmns-1,bib:ymois}]; this short scale associated with related
singular perturbation [first term of Eq. (\ref{DB5})] of
the system introduces the new channel for energy
transformations \citep{bib:mnsy,bib:msms,bib:BSM_deg,bib:SMB_multi}. If one
ignores this short scale (neglecting both Hall term and two-fluid
effects) one would end up creating the significant radial flows at
much larger distances from the stellar surface or the process of
flow acceleration will be completely eliminated from the dynamics
[see, for instance, the 2.5D-simulation results for magneto-fluid couplings
for classical solar case - Fig.12 of Mahajan et al 2006].

To better illustrate the coupling process we have picked up
several sets of our runs. The simulation results in Figures 1 and
2 for various physical quantities versus height consist of 3
frames (e.g Figs 1a and 1b; Figs 1c and 1d; Figs 1e and 1f), each
consisting of two pictures: one for the density and magnetic field
and the other for the velocity field. In each frame there are 3
sets of curves corresponding, respectively, to $\kappa = 10^{-2} ,
\ \kappa = 10^{-4} , \ \kappa = 10^{-6}$. Parameters defining
different frames are (we are giving them in the order \ $a=d, \
n_0,  \ B_0, \ V_A $) \ as follows: for Fig.1: (1) $10^{-1} ,
10^{26}\,cm^{-3} , 10^8\,G , 15\,km/s $ \ implying
\ $ a_0=(1/6)^{1/3}\,10^{-1}, \ V_0=10V_A=150\,km/s  $, \\
$\mu_0=10^7 $ \ (Figs.1a and 1b); (2) $ 10^{-1} , 10^{26}\,cm^{-3}
, 3\cdot 10^7\,G , 5\,km/s $ \ implying \
$a_0=(1/6)^{1/3}\,10^{-1}, \ V_0=10V_A=50\,km/s $, \
$ \mu_0=10^8 $ \ (Figs.1c and 1d); \\
(3) $ 10^{-1} , 10^{26}\,cm^{-3} ,$ \  $10^7\,G , 1.5\,km/s $ \
implying \ $ a_0=(1/6)^{1/3}\,10^{-1}, \ V_0=10V_A=15\,km/s, \
\mu_0=10^9 $ \ (Figs.1e and 1f) and for Fig.2: (4) $ 10 ,
10^{27}\,cm^{-3} , 3\cdot 10^8\,G , 15\,km/s $ \ implying \ $
a_0=(1/6)^{1/3}\,10^{-2/3}, \ V_0=0.1\,V_A=1.5\,km/s, \ \mu_0=10^7
$ \ (Figs.2a and 2b); \\
(5) $ 10 , 10^{27}\,cm^{-3} , 10^8\,G , 5\,km/s  $ \ implying \\
$ a_0=(1/6)^{1/3}\,10^{-2/3}, \ V_0=0.1\,V_A=0.5\,km/s, \
\mu_0=10^8 $ \ (Figs.2c and 2d); \ (6) $ 10 , 10^{27}\,cm^{-3} $,
\ $3\cdot 10^7\,G $, \\
$1.5\,km/s$ \ implying \ $ a_0=(1/6)^{1/3}\,10^{-2/3}, \
V_0=0.1\,V_A=0.15\,km/s, \ \mu_0=10^9 $ (Figs.2e and 2f).
\begin{figure}
\begin{center}
\includegraphics[scale=0.65,angle=0]{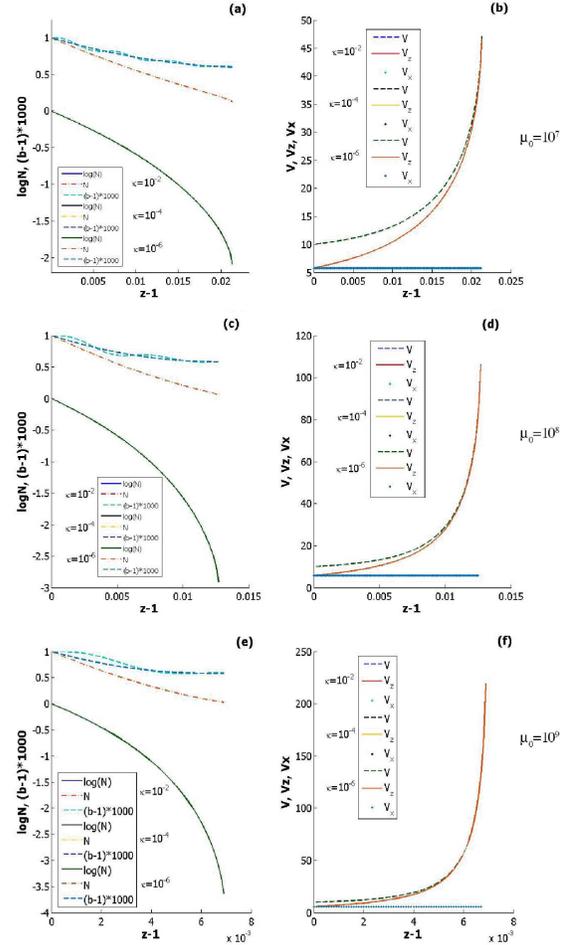}
\caption{ \ \ Plots for density, magnetic fields and velocity vs
height for $n_0=10^{26}\,cm^{-3} [a_0=(1/6)^{1/3}\,10^{-1}]$ and
DB parameter $a=d=0.1$ at various values of $\mu_0$ : (a) and (b)
are for $\mu_0=10^7$ and $R_A/\mu_0=0.05$; (c) and (d) are for
$\mu_0=10^8$ and $R_A/\mu_0=0.1$ and (e) and (f) are for
$\mu_0=10^9$ and $R_A/\mu_0=0.2$. The 3 sets of curves represent
results for $\kappa = 10^{-2} , \ \kappa = 10^{-4} , \kappa =
10^{-6}$, respectively. $V_y$ is not displayed due to its similar
behavior to $V_x$. For all $\kappa $-s one clearly sees the
dominance of radial velocity -- initial flow accelerates
significantly. Magnetic field components show oscillating
character although its energy doesn't change much at these heights
for chosen Beltrami parameters.} \label{Fig.1.}
\end{center}
\end{figure}

\begin{figure}
\begin{center}
\includegraphics[scale=0.8,angle=0]{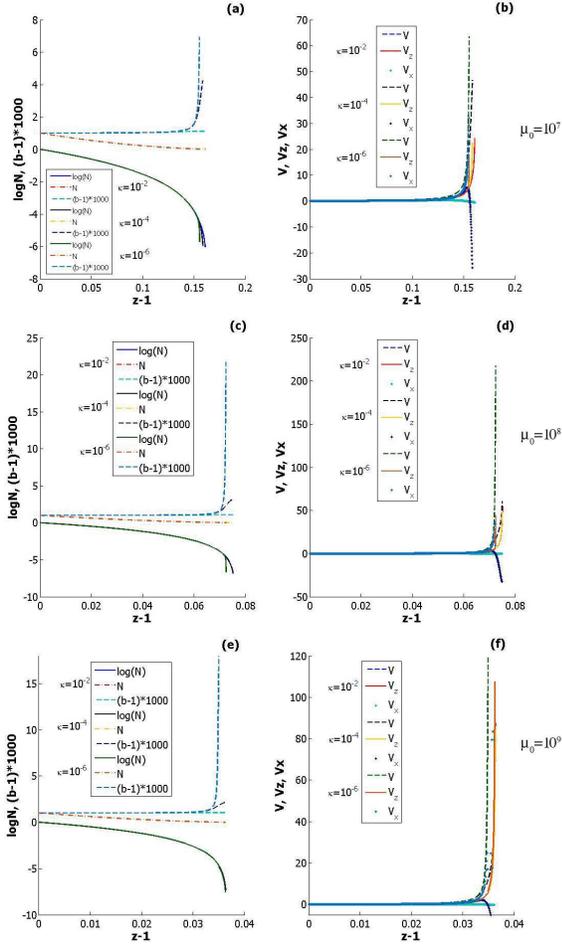}
\caption{ \ \ Plots for density, magnetic fields and velocity vs
height for $n_0=10^{27}\,cm^{-3} [a_0=(1/6)^{1/3}\,10^{-2/3}]$ and
DB parameter $a=d=10$ various values of $\mu_0$ : (a) and (b)
are for $\mu_0=10^7$ and $R_A/\mu_0=0.05$; (c) and (d) are for
$\mu_0=10^8$ and $R_A/\mu_0=0.1$ and (e) and (f) are for
$\mu_0=10^9$ and $R_A/\mu_0=0.2$. The 3 sets of curves represent
results for $\kappa = 10^{-2} , \ \kappa = 10^{-4} , \kappa =
10^{-6}$, respectively. $V_y$ is not displayed due to its similar
behavior to $V_x$. For all $\kappa $-s initial flow accelerates
significantly. There is a clear evidence of magnetic field
amplification together with flow acceleration.} \label{Fig.2.}
\end{center}
\end{figure}

In Fig.3 we give the simulation results for $n_0=10^{25}\,cm^{-3}
[a_0=(1/6)^{1/3}\,10^{-4/3}]$ for $B_0 = 3\cdot 10^6\,G  , \ a=d =
10$ \ (implying $V_A=1.5\,km/s , \ \mu_0 = 10^9$ )\ illustrating
that for such surface conditions starting from $0.15\,km/s$ speed
flow we arrive to $60\,km/s$ accelerated (400 times!) flow but at
lower densities of $\sim 10^{24}\,cm^{-3}$. We can clearly see
from this picture that in low-density degenerate case the
flow-generation picture practically repeats the classical case
results \citep{mnsy}; at the same time one can distinguish the
important difference -- there is a slight enhancement of the
magnetic field as well! Although this enhancement ($\sim 5\cdot
10^{3}\,G$) is small compared to background field its catastrophic
character of the increment introduces the short-scale in the
magnetic field structure which could trigger the severe
transformations in the initial equilibrium system (see e.g.
\citep{osym}). Also, later on, dynamically, such flow-magnetic
field region may become a boundary condition for a separate
coupling process leading to more complicated picture of flow
generation at the end.
\begin{figure}
\begin{center}
\includegraphics[scale=0.95,angle=0]{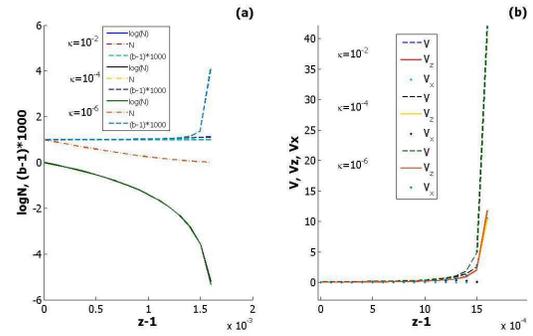}
\caption{ \ \ Plots for density, magnetic fields and velocity vs
height for $n_0=10^{25}\,cm^{-3} [a_0=(1/6)^{1/3}\,10^{-4/3}]$ and
DB parameter $a=d=10$ at $\mu_0=10^9$. Initial flow with
$0.15\,km/s$ speed accelerates 450 times reaching the speed of
$60\,km/s$; process is accompanied by the magnetic field
amplification.} \label{Fig.3.}
\end{center}
\end{figure}

The simulations show that, like in the solar case, for small
$\kappa $-s there exists some height at which the density begins
to drop dramatically (sometimes 400-500 times) with a
corresponding sharp rise in the flow speed and, often [for $a\sim
d <1$] in magnetic field as well. Effect is stronger for high
$\mu_0$ (initially high density and low magnetic field region)
degenerate plasmas. It is also obvious, that since the DB
structure scales are small compared to $R_W$ in outer layers /
atmosphere of the WD (where our model applies) the gravity
contribution, like in the solar case \citep{bib:mnsy,bib:msms}, determines
the radial distance in WD's outer layer over which the fast and
"catastrophic" acceleration of flow may appear due to the
magneto-fluid coupling. At the heights where the flows are
insignificant (at very short distances from the WD's surface)
gravity controls the stratification but as we approach the flow
"blow-up" distances (the flow becomes relatively strong) the
self-consistent magneto-Bernoulli processes (with degeneracy
effects included) take over and control the density (and hence the
velocity or/and magnetic field) stratification.

\bigskip

We have also verified the analytical prediction of \citep{bib:BSM_deg}
for maximal flow speed [achieved after acceleration] to be
super-Alfv\'enic . Interestingly enough, as numerical simulation
analysis showed, even for unrealistic Hall term strength parameter
$\kappa \sim 10^{-6}$ at specific boundary conditions we can see
the tendency of accelerated flow to become at the end locally
almost super-Alfv\'enic (starting from sub-Alfv\'enic); for
instance, initially sub-Alfv\'enic flow ($V_A=5\,km/s ; \ a=d=10$)
with speed $V_0=0.5\,km/s$ and density $\sim 10^{27}\,cm^{-3}$
accelerates sharply (2500 times) up to $1250\,km/s$ becoming
slightly super-Alfv\'enic locally at $Z-Z_0=0.07R_W$ and with
lower density $\leq 10^{23}\,cm^{-3}$ (Figs. 2b and 2c); it is
clear that accelerated flow remains strongly super-Alfv\'enic
locally when starting from being super-Alfv\'enic initially due to
severe drop in density -- see Fig.1). Thus, on can expect that
lowering $\kappa $ the chances of ending up with super-Alfve\'enic
accelerated flows increase (see Fig.7 and Fig.8); the higher is
the initial density/the lower the initial magnetic field better
could be the picture ($\mu_0$ goes up with density/inverce
magnetic field). In some cases the magnetic energy remained
unchanged (although the magnetic fields components would change
self-consistently, assuring the varying magnetization properties),
while in some cases (see e.g. Fig.2 and Fig.3) the magnetic field
energy was changed sharply; in few cases its change was wave-like
[like e.g. in Fig.1]. For DB parameter $a=10 \,(>1)$ one can
distinguish the enhancement of ${\bf B}$-field for all densities
$\geq 10^{25}\,cm^{-3}$, this is well seen in Fig.3. It is
evident, that the sharp decrease in density with a corresponding
sharp rise in the flow speed (and also in magnetic field) is the
expression of Bernoulli constraint imposed by the magneto-fluid
equilibrium condition. Notice, that we have neglected the thermal
pressure -- keeping it would not change the main results due to
the observationally evident very high degeneracy parameter $\mu_0$
in compact stellar outer layers, thus, controlling the equilibrium
state.

\begin{figure}
\begin{center}
\includegraphics[scale=0.95,angle=0]{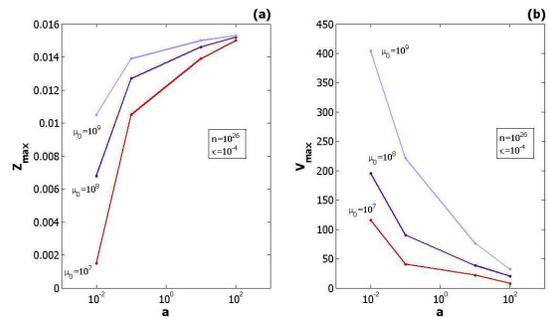}
\caption{ \ \ (a) Blow-up distance and (b) blow-up velocity versus
DB parameter $a\sim d$ for initial density $n_0=10^{26}\,cm^{-3}$.
The smaller the $\mu_0 $ lower is the final speed, as well as
smaller is the blow-up distance. For large $a$ both blow-up
distances and velocities come closer to each other for different
$\mu_0 $--s.} \label{Fig.4.}
\end{center}
\end{figure}

In Figures 4 and 5 we present the results for blow up distance
(left frame) and blow-up velocity (right frame) versus DB
parameter \ $a\sim d$ \ for different boundary conditions relevant
to Figs 1 and 2. We can clearly see that the final velocity
decreases with DB parameter and depends strongly on the degeneracy
parameter/magnetism of the stellar outer layer. The smaller is the
degeneracy parameter $\mu_0$ (low density/high magnetism) lower is
the final speed, blow-up distance is farther and vice-versa --
higher the degeneracy parameter $\mu_0$ (high density/low
magnetism) higher is the final speed for the same $a$ , blow-up
distance is closer to stellar surface. Also, from both figures one
can see, that for the same DB parameter $a \ll 1$ both {\it
Blow-up Distances} and {\it Blow-up Velocities} significantly
differ for different degeneracy parameter/magnetism (lower the DB
parameter more different they are) while for greater DB parameter
such difference disappears -- their values become similar for all
$\mu_0$ already for $a=100$. Hence, for large DB parameters both
{\it Blow-up Distance} and {\it Blow-up Velocity} are practically
independent of $a\sim d$. Such effect is stronger for larger
initial densities (see Fig.5).

\begin{figure}
\begin{center}
\includegraphics[scale=0.95,angle=0]{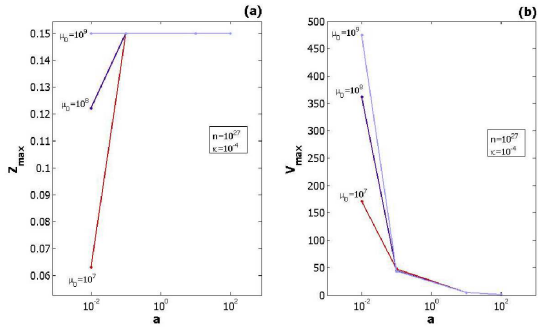}
\caption{ \ \ Blow-up (a) distance and (b) velocity versus DB
parameter $a\sim d$ for initial density $n_0=10^{27}\,cm^{-3}$.
For large $a>10^{-1}$ both blow-up distances and velocities become
practically independent of $a$ being similar for all $\mu_0 $--s.}
\label{Fig.2.}
\end{center}
\end{figure}

\begin{figure}
\begin{center}
\includegraphics[scale=1.10,angle=0]{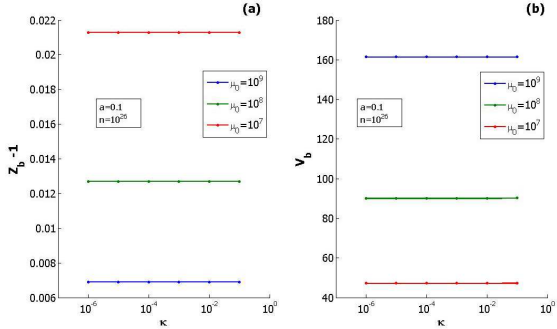}
\caption{ \ \ Blow-up (a) distance and (b) velocity vs Hall-term
strength $\kappa $ for $n_0=10^{26}\,cm^{-3}$ and $a=d=0.1$. The
smaller the $\mu_0 $ lower is the final speed and smaller is the
blow-up distance. For fixed $\mu_0 $ blow-up process is less
sensitive to changes in $\kappa $.} \label{Fig.6.}
\end{center}
\end{figure}

\begin{figure}
\begin{center}
\includegraphics[scale=1.10,angle=0]{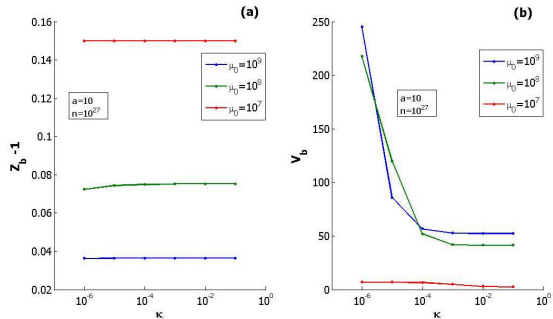}
\caption{ \ \ Blow-up (a) distance and (b) velocity vs $\kappa $
for $n_0=10^{27}\,cm^{-3}$ and $a=d=10$. For fixed $\mu_0 $
blow-up process is more sensitive to changes in $\kappa $ at
higher $\mu_0$ (lower initial magnetic field). There is a clear
tendency for initial flow to become Super-Alfv\'enic at blow-up.}
\label{Fig.7.}
\end{center}
\end{figure}
\begin{figure}
\begin{center}
\includegraphics[scale=1.10,angle=0]{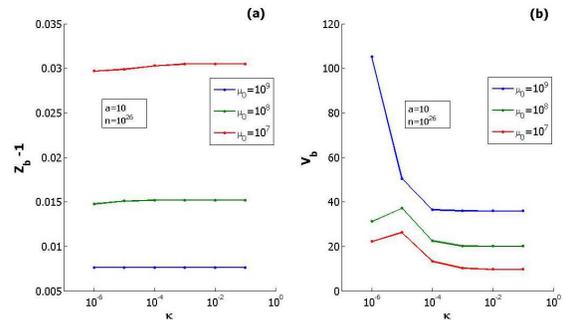}
\caption{ \ \ Blow-up (a) distance and (b) velocity vs $\kappa $
for $n_0=10^{26}\,cm^{-3}$ and 
$a\sim d=10$. } \label{Fig.8.}
\end{center}
\end{figure}


We have performed the detailed analysis to get more clear
information about the magneto-fluid coupling efficiency -- to
observe the dependence of flow/magnetic field blow-up process on
the Hall Current strength parameter $\kappa $. For larger $\kappa
$-s flows mostly tend to be radial, while for smaller $\kappa $-s
Double-Beltrami effects enter the play and the transverse
components of these vector-fields become more important. Situation
could change when we deal with the time-dependent dynamical model
with dissipation (e.g. when taking into account the
cooling/heating of the plasma); these processes would change
degeneracy parameter $\mu_0$ as well as the plasma $\beta_0 $
locally shifting the velocity/magnetic field blow-up distance at
lower/upper heights or eliminating it altogether -- one can
observe in Figs 1 and 2 that as $\mu_0$ goes up (density [magnetic
field] increases [decreases]) the fall in density [with
accompanying amplification of velocity/magnetic field] becomes
smoother. At fixed $\mu_0$ the blow-up distances and speeds are
not changed much with $\kappa$ for lower DB parameters ($a\sim d
<1$) (Fig.6), thus, indicating, that the efficiency of this
process is less sensitive to smallness rate of Hall Current
strength $\kappa $ (important is that it is not zero). At higher
$a\sim d \,[>1]$ for larger $\mu_0 $ one observes the increase of
amplification phenomenon at smaller $\kappa $-s -- see Figs 7 \&
8); effect is stronger for higher initial densities.


Notice that final velocities get larger with $V_0[km/s]$ \\
$\sim a^{-1}V_A$. For instance, initial flow with speed
$1.5\,km/s$ ends up acquiring a high speed $\sim 105\,km/s$ (Figs.
2a and 2b) at height $\sim 0.15R_W$ while initial flow
with speed $150\,km/s$ ends up acquiring a high speed $\sim
750\,km/s$ (Figs. 1a and 1b) at height $\sim 0.02R_W$
but at much lower densities compared to their initial one.

\bigskip

At this end we would like to summarize that there is an intrinsic
tendency of velocity / magnetic field amplification in our
two-fluid system of classical ions and degenerate elecrons due to
magneto-fluid coupling.  Also it is important to emphasize that
such coupling effect leads the system to the blow-up -- the
catastrophic amplification of magnetic/velocity vector-fields --
which can not be ignored in the dynamics of WD's outer layers with
degenerate electrons. Figs 4-8 clearly show, that the dependence
of both vector-fields (velocity and magnetic) during the blow-up
process on DB [$a ,\, d$] and degeneracy [$\mu_0 $] parameters is
fundamentally crucial: (1) for the different degeneracy parameters
the final results are different for the same $a\sim d$ (unless the
latter is much greater than one) while (2) for the same degeneracy
parameters final results may change considerably for different
$a$-s, thus, indicating that the blow-up process is very
sensitive, in fact, to both the degeneracy state of the system and
the magneto-fluid coupling. At the same time, since for greater DB
parameters the final blow-up speeds (blow-up distances) come
closer to each other for different $\mu_0 $-s it becomes
practically impossible to distinguish the realistic scenario
(select the corresponding boundary conditions) of
velocity/magnetic field amplification when comparing to
observations. If one ignores the flow term in Bernoulli constraint
(\ref{Bernoulli2}) one ends up finding the radial flows with very
low magnitude, and, like in Solar case \citep{bib:mnsy}, there will be
no region of sharp rise; also generated flows/magnetic fields will
achieve distinguishable energies on much longer distances from the
surface than the correct Bernoulli constraint would maintain it.
Thus, to find correctly the final values of amplified
vector-fields and also to determine the correct heights where such
catastrophic transformations would take place one has to perform
the magneto-fluid coupling analysis.

\section{Summary and Conclusions}


In present manuscript we extended the studies of recent papers
\citep{bib:BSM_deg,bib:SMB_multi} and, based on the systematic simulation
experiments, showed that the degeneracy effects are significant
for specific class of dense stellar atmospheres/outer layers
dynamics, specifically, for the structure formation phenomena
there -- we suggest that when studying the evolution of the
compact objects, flow effects cannot be ignored since their
catastrophic generation close to the surface may determine the
further evolution of stars and their atmospheres. In addition, as
observations indicate, for evolution picture, specifically in case
of magnetic WDs, it is of utmost importance to know the magnetic
fields dynamics/fate. In this respect our study indicates that for
specific WD-surface conditions it is possible to amplify the
magnetic fields in combination with flow generation/acceleration.
This simultaneous possibility of amplifying catastrophically flow
and magnetic fields is a demonstration of magneto-fluid coupling
-- here the degeneracy has a striking effect. The distance over
which amplification appears is determined by the strength of
gravity and degeneracy parameter; Double-Beltrami parameters also
play the crucial role in the efficiency of coupling. Examining the
concrete applications for cool WDs atmospheres we found the
appropriate physical parameter ranges for the flow
generation/acceleration and the magnetic field amplification;
process is less sensitive to surface temperature since the major
pressure is due to degeneracy. The possibility of locally
Super-Alfv\'enic flow generation is found for various surface
parameters of WDs -- this is a remarkable result compared to Sun
for which the generated flows (by the same mechanism) were
previously found to be locally sub-Alfv\'enic for low-beta plasmas
\citep{bib:mnsy}. For specific conditions of WD's outer layers initial
sub-Alfv\'enic flow can be accelerated 100 and more times leading
to the transient jet formation. The results of given study can be
applied for the understanding of the origin and evolution of White
Dwarfs, dense compact objects; for their cooling and accretion
dynamics for which, as demonstrated in present study, the
magneto-fluid coupling processes when degeneracy effects are taken
into account, specifically when time-dependency is also included,
become determining and crucial.

\section{Acknowledgements}

NLS acknowledges special debt to the Abdus Salam International
Centre for Theoretical Physics, Trieste, Italy. AAB thanks Doctors
A. Tevzadze and G. Mamatsashvili for the valuable discussions.


%



%

\end{document}